\documentclass[12pt,letterpaper]{article}
\pdfoutput=1
\usepackage{jheppub}
\usepackage{amsfonts, amsthm}
\usepackage[english]{babel}
\usepackage[utf8]{inputenc}
\usepackage{slashed}
\hypersetup{unicode}

\newcommand{\eq}{\begin{equation}}
\newcommand{\feq}{\end{equation}}
\newcommand{\eqn}{\begin{eqnarray}}
\newcommand{\feqn}{\end{eqnarray}}

\newcommand{\ma}[1]{\mbox{$\mathcal{#1}$}}

\title{Black holes in an expanding universe from fake supergravity}

\author{Samuele Chimento}
\author{and Dietmar Klemm}
\affiliation{Dipartimento di Fisica, Universit\`a di Milano, and \\
INFN, Sezione di Milano, \\
Via Celoria 16, 20133 Milano, Italy.
}
\emailAdd{samuele.chimento@mi.infn.it}
\emailAdd{dietmar.klemm@mi.infn.it}
\preprint{IFUM-1004-FT}
\abstract{In arXiv:0902.4814, a general recipe to construct fake supersymmetric solutions
to fake $N=2$, $d=4$ gauged supergravity coupled to abelian vector multiplets was presented.
We use these results to find new multi-centered black hole solutions in an asymptotically FLRW universe.
These satisfy the weak energy condition and are maximally charged under two $\text{U}(1)$ gauge fields
coupled to a scalar, which drives the cosmic expansion while rolling down its potential.
As a special subcase, our black holes include the ones constructed previously by Gibbons and Maeda
in arXiv:0912.2809. The latter contain two non-negative real numbers $n_S$, $n_T$ obeying the constraint
$n_S+n_T=4$, with the cases $n_T=4$ and $n_T=1$ corresponding to the Kastor-Traschen and
the Maeda-Ohta-Uzawa solution respectively. We show that $n_S$, $n_T$ arise directly as exponents
in the prepotential of the fake supergravity theory, and that the above constraint stems from the fact
that the prepotential must be a homogeneous function of degree two.
Finally, some physical properties of the black holes, like
asymptotic behaviour, curvature singularities and trapping horizons, are also discussed. Similar to other
solutions that appeared previously in the literature, there is a symmetry enhancement near the event horizon,
which becomes therefore a Killing horizon, in spite of the highly dynamical nature of the original
spacetime. The temperature associated to this Killing horizon turns out to be nonvanishing.
}

\keywords{Black Holes, Supergravity Models, Black Holes in String Theory.}

\begin{document}
\maketitle
\flushbottom

\section{Introduction}

Since the seventies of the last century, the physics of black holes has raised several fascinating
problems and puzzles, whose resolution is believed to be crucial for the construction of a future
quantum theory of gravity. Indeed, much of what we presently know on quantum effects in strong
gravitational fields comes from the study of stationary black holes, which by now are quite well
understood.

On the other hand, much less is known on dynamical processes involving black holes, since only a few
time-dependent black hole solutions have been constructed so far. The first and perhaps most famous
one is the McVittie spacetime \cite{McVittie:1933zz}, but there have been some controversies in the
literature concerning its interpretation as a black hole embedded in an FLRW
universe \cite{Nolan:1998xs,Nolan:1999kk,Kaloper:2010ec}. Using conformal techniques, Sultana and
Dyer \cite{Sultana:2005tp} constructed a black hole in a dynamical background, which however suffers
from the violation of energy conditions.

Other notable exceptions include the Kastor-Traschen solution \cite{Kastor:1992nn}, that describes
an arbitrary number of black holes in de~Sitter space, each of which carrying an electric charge
equal to the mass. This leads to a no-force condition, such that the whole system is just
comoving with the cosmological expansion. Five-dimensional multi-centered rotating charged
de~Sitter black holes were constructed in \cite{Klemm:2000vn,Klemm:2000gh}.
Maeda, Ohta and Uzawa (MOU in what follows) \cite{Maeda:2009zi} used
dynamical intersecting brane solutions in higher dimensions to obtain four- and five-dimensional
black holes that asymptotically tend to an FLRW universe filled with stiff matter.
These spacetimes were further studied in \cite{Maeda:2009ds}, and generalized to arbitrary dimension
in \cite{Maeda:2010ja}\footnote{See also \cite{Maeda:2011sh} for a review.}.
Generically, the solutions in \cite{Kastor:1992nn,Maeda:2009zi,Maeda:2010ja} are given in terms of
harmonic functions, such that one can superpose an arbitrary number of black holes, in spite of
the lack of supersymmetry. For the five-dimensional case of \cite{Maeda:2010ja}, it was shown in
\cite{Nozawa:2010zg} that this equilibrium condition can be traced back to the existence of a fake
Killing spinor\footnote{Kastor and Traschen showed in \cite{Kastor:1993mj} that their solution
satisfies the Killing spinor equation of minimal gauged supergravity, with a Wick-rotated coupling
constant, $g=iH$. In a more modern language, we would call this a fake Killing spinor equation.}.
By solving the first-order fake Killing spinor equations rather than the second-order Einstein
equations, the authors of \cite{Nozawa:2010zg} were then able to find a spinning generalization
of their five-dimensional cosmological black hole.

In this paper, we will pick up on this idea, in order to construct new multi-centered black hole
solutions in an asymptotically FLRW universe in four dimensions. To this end, we shall make
essential use of the results of \cite{Meessen:2009ma}, where a general recipe to construct fake
supersymmetric solutions to fake $N=2$, $d=4$ gauged supergravity coupled to (non)abelian vector
multiplets was provided\footnote{For related work in five dimensions cf.~\cite{Gutowski:2010sx}.}.
We shall choose models containing just one abelian vector multiplet, such that our solutions are charged
under two $\text{U}(1)$ gauge fields coupled to a scalar. As a special subcase, they include the black
holes constructed previously by Gibbons and Maeda \cite{Gibbons:2009dr}. The latter contain two
non-negative real numbers $n_S$, $n_T$ obeying the constraint $n_S+n_T=4$, with the cases
$n_T=4$ and $n_T=1$ corresponding to the Kastor-Traschen and the MOU solution respectively.
We show that $n_S$, $n_T$ arise as exponents in the prepotential of the fake supergravity
theory, and the constraint $n_S+n_T=4$ (which is somehow ad hoc in \cite{Gibbons:2009dr})
simply corresponds to the requirement that the prepotential must be a homogeneous function of
degree two. Moreover, our paper clarifies that the superposition principle in the Gibbons-Maeda
solution is just a consequence of the existence of a fake Killing spinor.
Although the matter content in the Lagrangian is the same, the multi-centered dynamical black holes
constructed here are more general than the ones in \cite{Gibbons:2009dr}; for instance our scalar potential
is a sum of several exponentials, which reduces to the Liouville-type potential of \cite{Gibbons:2009dr}
if one of the constants characterizing the model is set to zero.

The remainder of this paper is organized as follows: In section \ref{fake-sugra} we review fake
$N=2$, $d=4$ Fayet-Iliopoulos gauged supergravity, and the recipe to construct fake supersymmetric
solutions obtained in \cite{Meessen:2009ma}. In section \ref{sec:mod2}, we consider a simple prepotential
that contains one vector multiplet, construct dynamical multi-centered black hole solutions to this model,
and discuss in some detail their physical properties. In \ref{alternate-prepot}, a slightly different
model is considered, and it is shown that this leads to the Gibbons-Maeda spacetime, with
one of the two $\text{U}(1)$ gauge fields dualized. We conclude in section \ref{final} with some final remarks.

\section{Fake \texorpdfstring{$N=2$, $d=4$}{N=2, d=4} gauged supergravity}
\label{fake-sugra}

\subsection{Special geometry}
In $N=2$, $d=4$ supergravity coupled to $n_V$ vector multiplets, the complex scalars of the multiplets
parametrize an $n_V$-dimensional K\"ahler-Hodge manifold, which is the base of a symplectic bundle
with the covariantly holomorphic sections\footnote{Here and in what follows we use the conventions
of \cite{Meessen:2009ma}.}
\begin{equation}
 \mathcal{V}=\left(\begin{array}{c}
                   \mathcal{L}^\Lambda\\
                   \mathcal{M}_\Lambda
                  \end{array}\right), \qquad \mathcal{D}_{\bar \imath}\mathcal{V}\equiv\partial_{\bar \imath}\mathcal{V}-\frac{1}{2}\left(\partial_{\bar \imath}\mathcal{K}\right)\mathcal{V}=0\,,
\end{equation}
obeying the constraint
\begin{equation}
 \left\langle\mathcal{V},\bar{\mathcal{V}}\right\rangle\equiv\bar{\mathcal{L}}^\Lambda\mathcal{M}_\Lambda-\mathcal{L}^\Lambda\bar{\mathcal{M}}_\Lambda=-i\,, \label{eq:sympcond}
\end{equation}
where $\mathcal{K}$ is the K\"ahler potential.
We also introduce the explicitly holomorphic sections
\begin{equation}
 \Omega\equiv e^{-\mathcal{K}/2}\mathcal{V}\equiv\left(\begin{array}{c}
						  \mathcal{\chi}^\Lambda\\
						  \mathcal{F}_\Lambda
						  \end{array}\right);
\end{equation}
if the theory is defined by a prepotential $\mathcal{F}(\chi)$, then $\mathcal{F}_\Lambda=\partial_\Lambda \mathcal{F}$.
In terms of the sections $\Omega$ the constraint (\ref{eq:sympcond}) becomes
\begin{equation}
 \left\langle\Omega,\bar{\Omega}\right\rangle\equiv\bar{\chi}^\Lambda\mathcal{F}_\Lambda-\chi^\Lambda\bar{\mathcal{F}}_\Lambda=-i e^{-\mathcal{K}}.\label{eq:sympcond2}
\end{equation}
The couplings of the vector fields with the scalars are determined by the matrix $\mathcal{N}$, defined by the 
relations
\begin{equation}
 \mathcal{M}_\Lambda = \mathcal{N}_{\Lambda\Sigma}\,\mathcal{L}^\Sigma, 
 \qquad \mathcal{D}_{\bar \imath}\bar{\mathcal{M}}_\Lambda=\mathcal{N}_{\Lambda\Sigma}\,\mathcal{D}_{\bar \imath}\bar{\mathcal{L}}^\Sigma\,.
\end{equation}
In a theory with a prepotential, $\mathcal{N}$ can be obtained from
\begin{equation}
 \mathcal{N}_{\Lambda\Sigma}=\bar{\mathcal{F}}_{\Lambda\Sigma}+2i\frac{\mathfrak{Im}(\mathcal{F})_{\Lambda\Lambda'}\chi^{\Lambda'}\mathfrak{Im}(\mathcal{F})_{\Sigma\Sigma'}\chi^{\Sigma'}}{\chi^{\Omega}\mathfrak{Im}(\mathcal{F})_{\Omega\Omega'}\chi^{\Omega'}}\,, \label{eq:nmatrix}
\end{equation}
where $\mathcal{F}_{\Lambda\Sigma}=\partial_\Lambda\partial_\Sigma \mathcal{F}$.
% from this we can also get
% \begin{equation}
%  \mathfrak{Im}(\mathcal{N})^{-1|\Lambda\Sigma}=-\mathfrak{Im}(\mathcal{F})^{-1|\Lambda\Sigma}-2\mathcal{L}^\Lambda\bar{\mathcal{L}}^\Sigma-2\bar{\mathcal{L}}^\Lambda\mathcal{L}^\Sigma\label{eq:inverseimn}
% \end{equation}

The bosonic Lagrangian in the case of abelian vector multiplets, and with Fayet-Iliopoulos gauging of a
$\text{U}(1)$ R-symmetry subgroup, takes the form
\begin{align}
 e^{-1}\mathcal{L}_{\text{bos}}=&\,R+2\mathcal{G}_{i\bar\jmath}\partial_a Z^i\partial^a \bar Z^{\bar\jmath}-V
                                                          \nonumber\\
                         &+2\mathfrak{Im}(\mathcal{N})_{\Lambda\Sigma}F^\Lambda_{ab}F^{\Sigma ab}-2\mathfrak{Re}(\mathcal{N})_{\Lambda\Sigma}F^\Lambda_{ab}\star F^{\Sigma ab}\,,\label{eq:general_lagrangian}
\end{align}
where the gauging is realized with the connection $C_\Lambda A^\Lambda$ and the scalar potential is given by
\begin{equation}
 V=-\frac{g^2}{2}\left[4\left|C_\Lambda\mathcal{L}^\Lambda\right|^2+\frac{1}{2}\mathfrak{Im}(\mathcal{N})^{-1|\Lambda\Sigma}C_\Lambda C_\Sigma\right].\label{eq:scalar_pot}
\end{equation}
Since the matrix $\mathfrak{Im}(\mathcal{N})_{\Lambda\Sigma}$ appears in the kinetic term of the vector fields,
it must be negative definite and thus invertible. It can therefore be used as a `metric' to raise and lower
$\Lambda,\Sigma,\dots$ indices.

\subsection{Fake Killing spinors}

If we perform a Wick rotation on the gauge coupling constant of the theory, $g\rightarrow ig$, we obtain a new, 
non-supersymmetric theory with $V\rightarrow -V$ and a gauged $\mathbb{R}$-symmetry; the Killing spinor equations, coming 
from the vanishing of the fermionic supersymmetry variations, become
\begin{align}
 \mathbb{D}_a\epsilon_I&=\left[-2i\mathcal{L}_\Lambda F^{\Lambda +}_{ab}\gamma^b-\frac{ig}{4}C_\Lambda\mathcal{L}^\Lambda\gamma_a\right]\varepsilon_{IJ}\epsilon^J\,,\nonumber\\
 i\slashed{\partial}Z^i \epsilon^I&=\left[\bar f^i_\Lambda\slashed{F}^{\Lambda +}-\frac{g}{2}C_\Lambda\bar f^{i\Lambda}\right]\varepsilon^{IJ}\epsilon_J\,,
\end{align}
where
\[
\mathbb{D}_a\epsilon_I\equiv\left(\nabla_a +\frac{i}{2}\mathcal{Q}_a-\frac{g}{2}C_\Lambda
A_a^\Lambda\right)\epsilon_I\,,
\]
$\mathcal{Q}_a=(2i)^{-1}\left(\partial_a Z^i\partial_i\mathcal{K}-\partial_a \bar Z^{\bar\imath}\partial_{\bar\imath}\mathcal{K}\right)$ is 
the gauge field of the K\"ahler $\text{U}(1)$, and $f_i^\Lambda\equiv \mathcal{D}_i \mathcal{L}^\Lambda=\left(\partial_i+\frac{1}{2}\partial_i\mathcal{K}\right)\mathcal{L}^\Lambda$.

Since these equations do not come from supersymmetry, they are called fake Killing spinor equations, and solutions for which
they are satisfied are known as fake supersymmetric.

From the fake Killing spinors one can construct the bilinears
\begin{equation}
X=\frac{1}{2} \varepsilon^{IJ}\bar\epsilon_I\epsilon_J\,, \qquad
V_a=i\bar\epsilon^I\gamma_a\epsilon_I\,, \qquad
V_a^x=i(\sigma^x)_I^{\phantom{I}J}\bar\epsilon^I\gamma_a\epsilon_J\,,\label{eq:bilinears}
\end{equation}
and the real symplectic sections of K\"ahler weight zero
\begin{equation}
\mathcal{R}\equiv \mathfrak{Re}(\mathcal{V}/X)\,, \qquad
\mathcal{I}\equiv \mathfrak{Im}(\mathcal{V}/X)\,. \label{eq:risections}
\end{equation}

\subsection{Fake supersymmetric solutions}
In \cite{Meessen:2009ma}, Meessen and Palomo-Lozano presented a general method to obtain fake
supersymmetric solutions to Wick-rotated $N=2$, $d=4$ gauged supergravity coupled to nonabelian vector
multiplets. We will restrict ourselves here to the case of just abelian multiplets, with no gaugings of the scalar
manifold's isometries; we will also consider only the timelike case of \cite{Meessen:2009ma}, which means
that we take the norm of $V$ defined in (\ref{eq:bilinears}) to be positive.

With these restrictions, the fake supersymmetric solutions always assume the form
\begin{align}
 ds^2&=2\left|X\right|^2(d\tau+\omega)^2-\frac{1}{2\left|X\right|^2}h_{mn}dy^mdy^n\,, \\
 A^\Lambda&=-\frac{1}{2}\mathcal{R}^\Lambda V + \tilde{A}^\Lambda_m dy^m\,, \\
 Z^\Lambda&=\frac{\mathcal{L}^\Lambda}{\mathcal{L}^0}=\frac{\mathcal{R}^\Lambda+i\mathcal{I}^\Lambda}{\mathcal{R}^0+i\mathcal{I}^0}\,,
\end{align}
where $V=2\sqrt{2}\left|X\right|^2(d\tau+\omega)$, $\omega=\omega_m dy^m$ is a 1-form which can in
general depend on $\tau$, and $h$ is the metric on a three-dimensional Gauduchon-Tod \cite{gt}
base space. In particular there must exist a dreibein $W^x$ for $h$ satisfying
\begin{equation}
 dW^x=gC_\Lambda \tilde{A}^\Lambda \wedge W^x +\frac{g}{2\sqrt{2}} C_\Lambda {\mathcal{I}}^\Lambda \varepsilon^{xyz} W^y\wedge W^z.\label{eq:gt}
\end{equation}
Furthermore the following equations must hold:
\begin{align}
 &\omega=gC_\Lambda\tilde A^\Lambda\tau+\tilde\omega\,, \label{eq:omegataudep}\\
 &\tilde{F}^\Lambda_{xy}=-\frac{1}{\sqrt{2}}\varepsilon^{xyz}\tilde{\mathbb{D}}_z \mathcal{I}^\Lambda\,,
 \label{eq:bogo}\\
 &\partial_\tau \mathcal{I}^\Lambda=0\,, \qquad \partial_\tau \mathcal{I}_\Lambda=-\frac g{2\sqrt2}
 C_\Lambda\,, \label{eq:dertaui}\\
 &\tilde{\mathbb{D}}_x^2\tilde{\mathcal{I}}_\Lambda-\left(\tilde{\mathbb{D}}_x \tilde \omega_x\right)
 \partial_\tau\mathcal{I}_\Lambda =0\,, \label{eq:omegai1}\\
 &\tilde{\mathbb{D}}\,\tilde \omega=\frac12\varepsilon^{xyz}\left\langle\tilde{\mathcal{I}}\right.\left|\partial_x
 \tilde{\mathcal{I}}-\tilde\omega_x\partial_\tau\mathcal{I}\right\rangle W^y\wedge W^z\,, \label{eq:omegai2}
\end{align}
with
\begin{gather}
\tilde{F}^\Lambda\equiv d \tilde{A}^\Lambda\,, \qquad \tilde\omega\equiv\omega|_{\tau=0}\,, \qquad
\tilde{\mathcal{I}}\equiv\mathcal{I}|_{\tau=0}\,, \\
\tilde{\mathbb{D}}_m \mathcal{I}\equiv \partial_m\mathcal{I}+gC_\Lambda\tilde A^\Lambda_m\mathcal{I}\,,
\qquad \tilde{\mathbb{D}}_x \mathcal{I}\equiv W_x^m\tilde{\mathbb{D}}_m \mathcal{I}\,.
\end{gather}
To obtain a specific solution we will then have to take the following steps:
\begin{enumerate}
 \item Choose the number of vector multiplets, the real constants $C_\Lambda$ and the special geometric
 manifold, e.g. by specifying a prepotential; this completely determines the bosonic action and permits to
 derive the dependence of the 
 $\mathcal{R}$'s from the $\mathcal{I}$'s, the so-called \emph{stabilization equations}.
 \item Choose a three-dimensional Gauduchon-Tod base space, that is, choose a solution 
 $(W^x,C_\Lambda\tilde A^\Lambda,C_\Lambda \mathcal{I}^\Lambda)$ of equation (\ref{eq:gt}).
 \item Determine the $\mathcal{I}^\Lambda$'s and the $\tilde{A}^\Lambda$'s that respect the choices of points
 1 and 2 and at the same time satisfy equation (\ref{eq:bogo}).
 \item Determine the $\mathcal{I}_\Lambda$'s and $\tilde\omega$ from (\ref{eq:dertaui}) and the coupled
  equations (\ref{eq:omegai1}) and (\ref{eq:omegai2}).
 \item Solve the stabilization equations to find the $\mathcal{R}$'s and finally write down the metric and the
 fields of the solution using (\ref{eq:omegataudep}) and
 $1/\left|X\right|^2=2\left\langle\mathcal{R}|\mathcal{I}\right\rangle$.
\end{enumerate}

In the next sections, we will use this procedure to find some solutions to theories with one vector multiplet, so
that there will be only one physical scalar $Z^1\equiv Z$. We shall also make the simplest possible choice for
the Gauduchon-Tod base space, the flat space.

\section{The \texorpdfstring{$\mathcal{F}(\chi)=-\frac i4(\chi^0)^n(\chi^1)^{2-n}$}{ℱ(χ)=-i/4 (χ⁰)ⁿ (χ¹)²⁻ⁿ} model\label{sec:mod2}}

Given this prepotential with $n\neq 0,2$, from (\ref{eq:sympcond2}) we can derive the K\"ahler potential
\begin{equation}
 e^{-\mathcal{K}}=\frac{n}{4} Z^{2-n} +\frac{2-n}{4} \bar Z Z^{1-n} + \text{c.c.}\,,
\end{equation}
where we took $\left|\chi^0\right|=1$. 

If we consider the truncation $\mathfrak{Im}(Z)=0$, the K\"ahler metric becomes
\begin{equation}
 \mathcal{G}=\left.\partial_Z\partial_{\bar Z}\mathcal{K}\right|_{\mathfrak{Im}(Z)=0}=\frac{n(2-n)}4
 \mathfrak{Re}(Z)^{-2}=\frac{n_0n_1}{16}e^{-2\phi}\,,
\end{equation}
where we defined $n_0\equiv 2n$, $n_1\equiv 2(2-n)=4-n_0$, $\phi\equiv\log\mathfrak{Re}(Z)$.

From equation (\ref{eq:nmatrix}) we obtain then
\begin{equation}
 \mathcal{N}=-\frac{i}{8}\left(\begin{array}{cc}
                    n_0 e^{\frac{n_1}{2}\phi} & 0                 \\
                    0 & n_1 e^{-\frac{n_0}{2}\phi}
                   \end{array}
\right),
\end{equation}
and for the scalar potential (\ref{eq:scalar_pot}) we get
\begin{equation}
 V=\frac{1}{2}\left[\frac{n_0(n_0-1)}{t_0^2}e^{-\frac{n_1}{2}\phi}+2\frac{n_0n_1}{t_0t_1}e^{\frac{n_0-n_1}{4}\phi}
 +\frac{n_1(n_1-1)}{t_1^2}e^{\frac{n_0}{2}\phi}\right]\,,
\end{equation}
with the definition $t_\Lambda\equiv-\frac{n_\Lambda}{2gC_\Lambda}$. If one wishes to have a non-zero
potential in the particular cases $n_0=1$ and $n_0=3$ one has to require respectively $C_1\neq 0$ and
$C_0\neq 0$.

Plugging these expressions into (\ref{eq:general_lagrangian}) leads to the bosonic Lagrangian
\begin{align}
 e^{-1}\mathcal{L}=&R+\frac{n_0n_1}{8}\partial_\mu\phi\partial^\mu\phi-\frac{n_0}{4}e^{\frac{n_1}{2}\phi} F^0_{\mu\nu}F^{0\mu\nu}-\frac{n_1}{4}e^{-\frac{n_0}{2}\phi} F^1_{\mu\nu}F^{1\mu\nu}\nonumber\\
                   &-\frac{1}{2}\left[\frac{n_0(n_0-1)}{t_0^2}e^{-\frac{n_1}{2}\phi}+2\frac{n_0n_1}{t_0t_1}e^{\frac{n_0-n_1}{4}\phi}+\frac{n_1(n_1-1)}{t_1^2}e^{\frac{n_0}{2}\phi}\right]\,. \label{L-1stmodel}
\end{align}
We see that in order to avoid ghost fields in the Lagrangian one has to impose $0<n_0<4$, corresponding
to $0<n<2$ in the prepotential. One can check that for $t_1\to\infty$ (i.e., $C_1=0$), \eqref{L-1stmodel}
reduces to the Lagrangian used in \cite{Gibbons:2009dr}, if we identify $n_0=n_T$, $n_1=n_S$.

\eqref{eq:risections}, together with $\mathfrak{Im}(Z)=0$, leads to
\begin{gather}
 \mathcal{I}^1=e^\phi \mathcal{I}^0\,, \qquad \mathcal{I}_0=\frac{n_0}{n_1}e^\phi \mathcal{I}_1\,, \nonumber\\
 \mathcal{R}^0=-\frac{8}{n_1}e^{\frac{n_0-n_1}{4}\phi}\mathcal{I}_1\,, \qquad \mathcal{R}^1=-\frac8{n_1}
 e^{\frac{n_0}{2}\phi} \mathcal{I}_1\,, \nonumber\\
 \mathcal{R}_0=\frac{n_0}{8}e^{\frac{n_1}{2}\phi}\mathcal{I}^0\,, \qquad \mathcal{R}_1=\frac{n_1}8
 e^{\frac{n_1-n_0}{4}}\mathcal{I}^0\,, \label{eq:mod2ir}
\end{gather}
as well as
\begin{equation}
 \frac{1}{2|X|^2}=\langle\mathcal{R}|\mathcal{I}\rangle=\frac{1}{2}\,e^{\frac{n_1}{2}\phi}(\mathcal{I}^0)^2+\frac{32}{n_1^2}\,e^{\frac{n_0}{2}\phi}(\mathcal{I}_1)^2\,. \label{eq:mod2sympprod}
\end{equation}
Notice that since both $\mathcal{I}^0$ and $\mathcal{I}^1$ must be independent of $\tau$, either
$\mathcal{I}^0=0$ or $\phi$ is also independent of $\tau$. In this second case using (\ref{eq:dertaui}) we
see that $C_0=0 \Leftrightarrow C_1=0$, so that if we require a non vanishing scalar potential we must
impose $C_0,C_1\neq 0$; we also find that $e^\phi=\frac{n_1}{n_0}\frac{C_0}{C_1}=\frac{t_1}{t_0}$.

\subsection{Construction of the solution}

The simplest solution of eq. (\ref{eq:gt}) is the flat three-dimensional space, with
\begin{equation}
 W^x_m=\delta^x_m,\qquad C_\Lambda\tilde A^\Lambda=C_\Lambda \mathcal{I}^\Lambda=0\,.
\end{equation}
With this choice for the base space we don't need to distinguish between $x,y,z\dots$ and lower
$m,n,p,\dots$ indices. 

If we require $V\neq 0$, $C_\Lambda \mathcal{I}^\Lambda=0$ together with (\ref{eq:mod2ir}) implies either
\begin{equation}
 \mathcal{I}^0=0\qquad\Rightarrow\qquad \mathcal{I}^1=\mathcal{R}_0=\mathcal{R}_1=0\,, \label{eq:mod2flatzerosections}
\end{equation}
or a constant $\phi$ with
\begin{equation}
 e^\phi=-\frac{C_0}{C_1}\,,
\end{equation}
and $C_0, C_1\neq 0$; but if $\phi$ is constant we should also have $e^\phi=\frac{n_1}{n_0}\frac{C_0}{C_1}$,
so this choice is clearly inconsistent. The only consistent possibility is then $\mathcal{I}^0=0$. Using equation
(\ref{eq:bogo}) this immediately implies
\begin{equation}
 \tilde{F}^0=\tilde{F}^1=0\,.
\end{equation}

Because of (\ref{eq:mod2flatzerosections}) and $C_\Lambda \tilde{A}^\Lambda=0$, eq. (\ref{eq:omegai2})
implies $d\tilde\omega=0$, and thus locally $\tilde\omega=df$, where $f$ is a generic function of the spatial
coordinates. 

Equation (\ref{eq:omegai1}) then becomes
\begin{equation}
 \left\{\begin{array}{l}
	 \partial_p\partial_p(\tilde{\mathcal{I}}_0+\frac{g C_0}{2\sqrt{2}}f)=0\,, \\
         \partial_p\partial_p(\tilde{\mathcal{I}}_1+\frac{g C_1}{2\sqrt{2}}f)=0\,,
        \end{array}\right.
        \qquad\Rightarrow\qquad
 \left\{\begin{array}{l}
	 \partial_p\partial_p(e^{\tilde \phi}\tilde{\mathcal{I}}_1-\frac{n_1}{4\sqrt{2}}\frac{f}{t_0})=0\,, \\
         \partial_p\partial_p(\tilde{\mathcal{I}}_1-\frac{n_1}{4\sqrt{2}}\frac{f}{t_1})=0\,,
        \end{array}\right.
\end{equation}
with $\tilde\phi\equiv\phi|_{\tau=0}$. This can be solved by introducing two generic harmonic functions
of the spatial coordinates $\mathcal{H}_0, \mathcal{H}_1$ as
\begin{equation}
 \tilde{\mathcal{I}}_1=\frac{n_1}{4\sqrt{2}}(f/t_1+\mathcal{H}_1)\,, \qquad
 e^{\tilde\phi}=\frac{f/t_0+\mathcal{H}_0}{f/t_1+\mathcal{H}_1}\,.
\end{equation}
At this point, using (\ref{eq:dertaui}) and $\mathcal{I}_0=e^\phi \mathcal{I}_1$ we obtain
\begin{gather}
 \mathcal{I}_1=\frac{n_1}{4\sqrt{2}}\left(\frac{\tau+f}{t_1} + \mathcal{H}_1\right),\qquad \mathcal{I}_0=\frac{n_0}{4\sqrt{2}}\left(\frac{\tau+f}{t_0} + \mathcal{H}_0\right), \nonumber\\
 e^\phi=\frac{(\tau+f)/t_0 + \mathcal{H}_0}{(\tau+f)/t_1 + \mathcal{H}_1}\,, \label{eq:mod2_phi_tau_dep}
\end{gather}
and from (\ref{eq:mod2sympprod}) one gets
\begin{equation}
 \frac{1}{2|X|^2}=\left(\frac{\tau+f}{t_0} + \mathcal{H}_0\right)^\frac{n_0}{2}\left(\frac{\tau+f}{t_1} + \mathcal{H}_1\right)^\frac{n_1}{2}.
\end{equation}
We have now all the elements needed to write down the complete solution in terms of the two generic 
harmonic functions $\mathcal{H}_0$ and $\mathcal{H}_1$. Since $f$ appears everywhere as a shift in the
time coordinate $\tau$ we can set it equal to zero with the coordinate change $t=\tau+f$ to obtain
\begin{gather}
ds^2=\mathcal{U}^{-2}dt^2- \mathcal{U}^2 d\vec y^{\,2}\,, \label{eq:mod2_flat_sol} \\
A^{\Lambda}=\left(\frac{t}{t_{\Lambda}} + \mathcal{H}_{\Lambda}\right)^{-1}dt\,, \qquad
\phi=\ln\left(\frac{t/t_0+\mathcal{H}_0}{t/t_1+\mathcal{H}_1}\right)\,, \nonumber
\end{gather}
with
\begin{equation}
 \mathcal{U}\equiv\left(\frac{t}{t_0} + \mathcal{H}_0\right)^\frac{n_0}4\left(\frac{t}{t_1} + \mathcal{H}_1
 \right)^\frac{n_1}4\,.
\end{equation}
Here one clearly recognizes the substitution principle originally put forward by Behrndt and
Cveti\v{c} in \cite{Behrndt:2003cx}, which amounts to adding a linear time dependence to the harmonic
functions in a supersymmetric solution of $N=2$, $d=4$ supergravity.

\subsection{Physical discussion}

As a first remark if we set $C_1=0$, corresponding to $t_1\rightarrow \infty$, and make the choice of harmonic functions
\begin{equation}
 \mathcal{H}_0=\sum_{i=1}^N \frac{Q_0^{(i)}}{|\vec{y}-\vec{y}_i|}\,, \qquad \mathcal{H}_1=1+\sum_{i=1}^N
 \frac{Q_1^{(i)}}{|\vec{y}-\vec{y}_i|}\,,
\end{equation}
we recover precisely the solution presented in \cite{Gibbons:2009dr}. The same is true if we set $C_0=0$,
change the sign of the scalar field and exchange everywhere $0$ and $1$ indices. This solution represents a
system of multiple maximally charged black holes in a universe expanding with arbitrary equation of state
$P=w \rho$, with $w=\frac{8-5n_0}{3n_0}$ so  that $-1\leq w\leq 1$ for $1\leq n_0\leq 4$ (for $n_0<1$ the
scalar potential is unbounded from below). Note that one can have $w<-1$ by allowing $n_0<0$ or
$n_0>4$, but then of course the action \eqref{L-1stmodel} contains ghosts. In this case, we would have
black holes embedded in an expanding universe filled with phantom energy.
In the limit $n_0=4$ one obtains the Kastor-Traschen
solution \cite{Kastor:1992nn}, describing multiple black holes in a de~Sitter background, while for $n_0=0$
the scalar potential is zero and the solution is the 
Majumdar-Papapetrou spacetime, describing multiple extremal Reissner-Nordstr\"om black holes in an
asymptotically flat background. Notice that we
can also recover the Kastor-Traschen solution keeping both $t_0$ and $t_1$ finite and taking 
$t_0\mathcal{H}_0=t_1\mathcal{H}_1$.

Retaining both $C_0$ and $C_1$ the scalar potential has critical points; the derivative of the scalar potential
can be written as
\begin{equation}
 V'[\phi]=\frac{n_0 n_1}{4 t_1^2}e^{-\frac{n_1}{4}\phi}\left[\frac{t_1}{t_0}-e^\phi\right]\left[\frac{t_1}{t_0}(1-n_0)+(1-n_1)e^\phi\right].
\end{equation}
We can see that if we take $t_0 t_1 >0$ there is, for every value of $0<n_0<4$, a minimum in 
$e^\phi=\frac{t_1}{t_0}$, $V_{\text{min}}=6 \,t_1^{-n_1/2} t_0^{-n_0/2}$. For $0<n_0<1$ or $3<n_0<4$ there is also a maximum in $e^\phi=-\frac{1-n_0}{1-n_1}\frac{t_1}{t_0}$, 
$V_{\text{max}}=2\,(\frac{n_0-1}{1-n_1})^{\frac{n_0-n_1}{4}}t_1^{-n_1/2} t_0^{-n_0/2}$; however for these
values of $n_0$ the potential is not bounded from below. For $1<n_0<3$ the potential is bounded and the
minimum is global.

If on the other hand we take $t_0 t_1 <0$, there is only a negative minimum in $e^\phi=-\frac{1-n_0}{1-n_1}\frac{t_1}{t_0}$
if $1<n_0<3$, $V_{\text{min}}=-2\,(\frac{1-n_0}{1-n_1})^{\frac{n_0-n_1}{4}}|t_1|^{-n_1/2} |t_0|^{-n_0/2}$, while there are no critical points for $0<n_0<1$ or $3<n_0<4$.

For $t_0 t_1 >0$ and assuming that the harmonic functions have a well-defined limit for 
$|\vec{y}|\rightarrow\infty$, one can study the asymptotic behaviour of the metric; swapping the coordinate
$t$ for $\tilde t$ defined by
\begin{equation}
 \frac{d\tilde t}{dt}=\left(\frac{t}{t_0} + k_0\right)^{-\frac{n_0}4}\left(\frac{t}{t_1} + k_1\right)^{-\frac{n_1}4},
 \qquad k_i\equiv\lim_{|\vec{y}|\rightarrow\infty}\mathcal{H}_i\,,
\end{equation}
the metric asymptotically assumes a Friedmann-Lema\^itre-Robertson-Walker form, 
\begin{equation}
 ds^2=d\tilde t^2-a^2(\tilde t)d{\vec{y}}^2\,, \qquad a(\tilde t)=\frac{dt}{d\tilde t}\,. 
\end{equation}
The explicit form of $a(\tilde t)$ is complicated; however it is possible to obtain the time-dependence of
the density and pressure,
\begin{align}
 \rho(\tilde t)&=\frac{3}{128\pi} \frac{n_0^2}{t_0^2} \frac{1}{R(\tilde t)^{\frac{n_0}{2}}}\! \left( R(\tilde t) +\frac{n_1}{n_0}\frac{t_0}{t_1} \right)^{\!\!2}\,, \\
 P(\tilde t)   &=-\frac{5}{128\pi} \frac{n_0^2}{t_0^2} \frac{1}{R(\tilde t)^{\frac{n_0}{2}}}\!\left[\left( R(\tilde t) +\frac{n_1}{n_0}\frac{t_0}{t_1} \right)^{\!\!2}\!-\frac8{5 n_0}\left( R^2(\tilde t)+\frac{n_1}{n_0}\frac{t_0^2}{t_1^2} \right)\right]\,,
\end{align}
where
\begin{equation}
 R(\tilde t)\equiv\frac{t(\tilde t)/t_1+k_1}{t(\tilde t)/t_0+k_0}\,,
\end{equation}
so that
\begin{equation}
 \frac{P(\tilde t)}{\rho(\tilde t)} = w(\tilde t)=-\frac{5}{3}\left[1-\frac{8}{5n_0}\frac{R^2(\tilde t)+\frac{n_1}{n_0}\frac{t_0^2}{t_1^2}}{\left(R(\tilde t)+\frac{n_1}{n_0}\frac{t_0}{t_1}\right)^2}\right]\,,
\end{equation}
that gives the correct value of \cite{Gibbons:2009dr} in the limits $t_0\rightarrow\infty$ or $t_1\rightarrow\infty$. 

If both $t_0$ and $t_1$ are finite, $w$ is time-independent only if $t_0 k_0 = t_1 k_1$, which is equivalent
to consider $k_0=k_1=0$, since we are free to set $k_0=0$ without loss of generality by shifting $t$.
In this case $a(\tilde t)= e^{\tilde t/\tilde t_0}$, with $\tilde t_0=t_0^{n_0/4}t_1^{n_1/4}$, $w=-1$ 
and the spacetime is asymptotically de Sitter independently of the value of $n_0$, while the scalar field tends to the critical value $e^\phi=t_1/t_0$.

Note that in the case $t_0t_1>0$, the solution \eqref{eq:mod2_flat_sol} tends to de~Sitter for
$|\vec y|\to\infty$ and arbitrary $k_{\Lambda}$ either for $t\to\infty$ or $t\to-\infty$ (for positive or negative
$t_{\Lambda}$ respectively).

Since we are interested in black hole systems, we consider harmonic functions of the form
\begin{equation}
 H_{\Lambda}(t,\vec{y})\equiv\frac{t}{t_{\Lambda}}+\mathcal{H}_{\Lambda}=\frac{t}{t_{\Lambda}} +
 k_{\Lambda}+\sum_{i=1}^{N} \frac{Q^{(i)}_{\Lambda}}{|\vec{y}-\vec{y}_i|}\,,
\end{equation}
and take $k_0=0$ since it can be eliminated by shifting $t$. Notice that while we could take some of the 
charges to be zero, this would lead to a divergent scalar field in the limit $|\vec{y}-\vec{y}_i|\rightarrow 0$.

The scalar curvature of \eqref{eq:mod2_flat_sol} reads
\begin{align}
 R=&\frac{3}{8}\frac{n_1(3n_1-4)t_0^2 H_0^2+6n_0n_1t_0t_1H_0H_1+n_0(3n_0-4)t_1^2H_1^2}{t_0^2t_1^2H_0^\frac{n_1}2 H_1^{\frac{n_0}2}}\nonumber\\
   &+\frac{n_0n_1}{8 H_0^{\frac{n_0}2}H_1^{\frac{n_1}{2}}}\left(\partial_p\ln\frac{H_0}{H_1}\right)^2\,,
\end{align}
which is singular for $H_0=0$ or $H_1=0$.

We can also consider the limit $|\vec{y}-\vec{y}_i|\equiv r_i\rightarrow 0$ for some $i$; then the time
dependence drops out and the metric reduces to $\text{AdS}_2\times\text{S}^2$,
\begin{equation}\label{throat}
 ds^2_{r_i\rightarrow 0}=\frac{r_i^2}{l_i^2}dt^2-\frac{l_i^2}{r_i^2}dr_i^2-l_i^2 d\Omega_2^2\,,
\end{equation}
with $l_i\equiv(Q_0^{(i)})^{n_0/4}(Q_1^{(i)})^{n_1/4}$. As we shall see later, \eqref{throat} does actually
not describe the geometry near the event horizon of our time-dependent solution.

We turn now to study in more detail the system with a single black hole. Since in this case there is spherical symmetry, we will work in spherical coordinates,
\begin{equation}
 H_0(t,r)=\frac{t}{t_0} + \frac{Q_0}{r}\,, \qquad H_1(t,r)=\frac{t}{t_1} + k_1 + \frac{Q_1}{r}\,.
\end{equation}
If $Q_0,\,Q_1\neq 0$ we will assume in the following, without loss of generality, $|Q_1t_1|\ge|Q_0t_0|$.

\begin{figure}[htb]
 \begin{center}
  \includegraphics[scale=0.1]{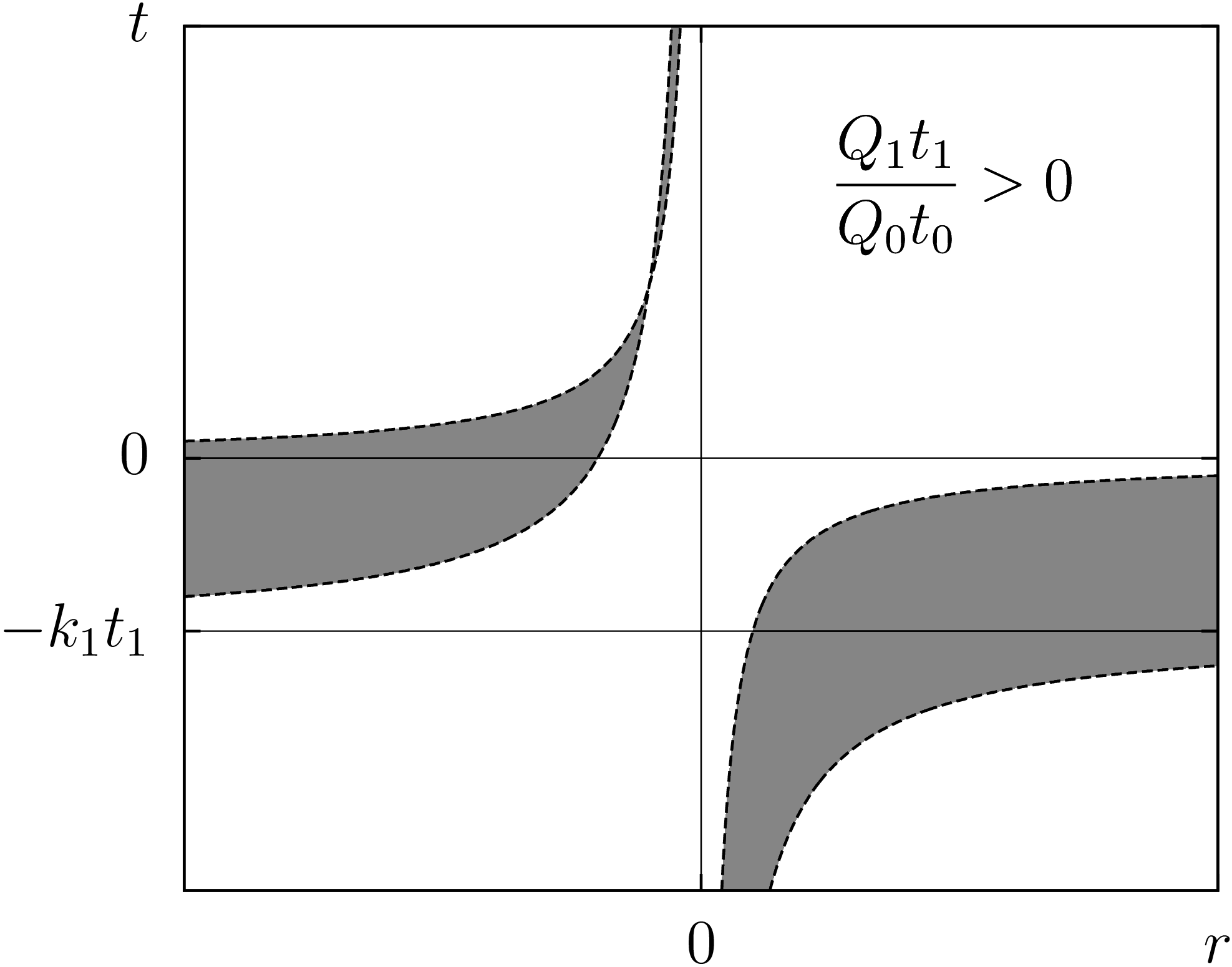}\hspace{0.5cm}\includegraphics[scale=0.1]{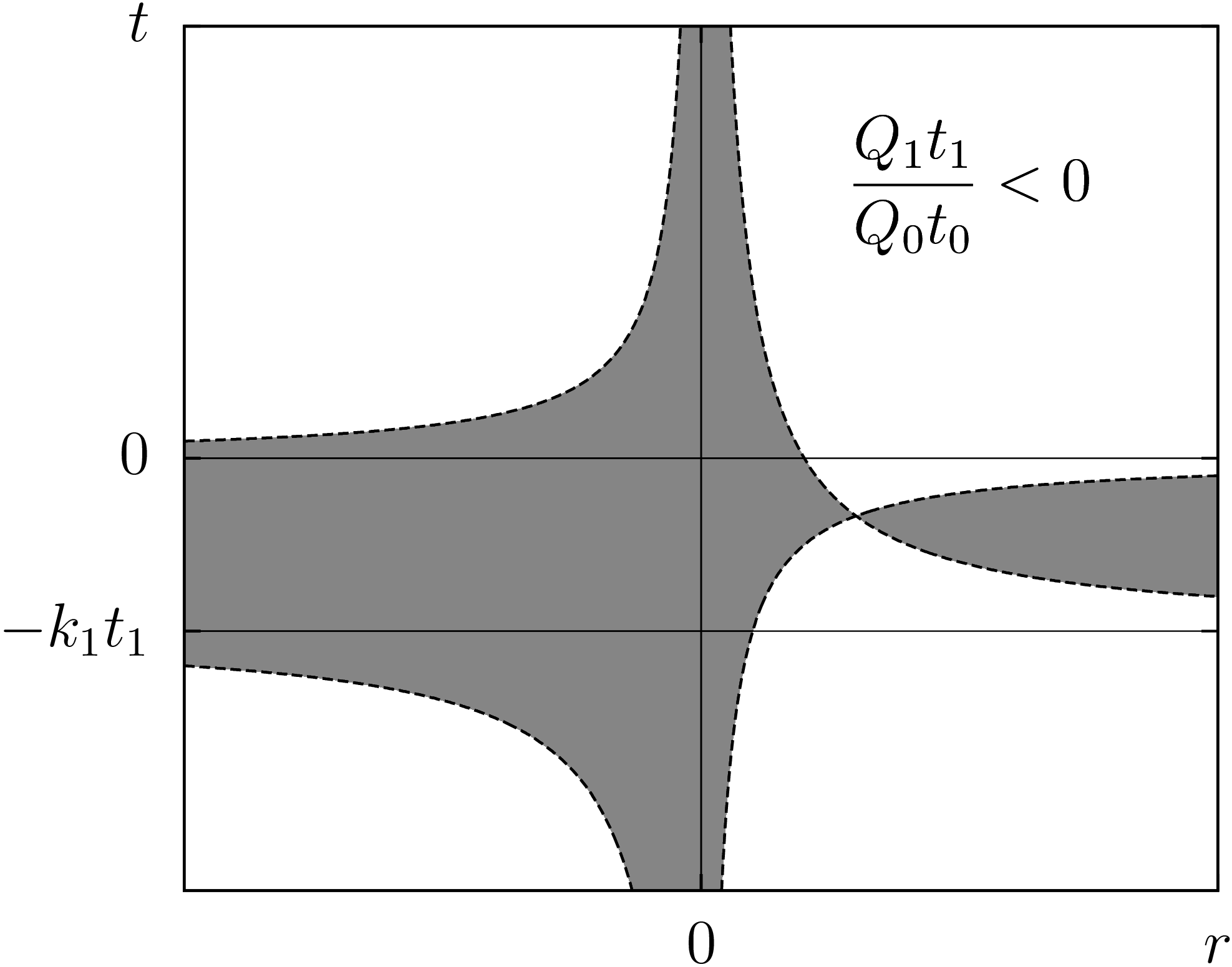}
 \end{center}
 \caption{Allowed coordinate ranges in the $r-t$ plane. The dashed curves denote the curvature singularities, the allowed 
 range is the white area for $t_0t_1>0$ or the grey area for $t_0t_1<0$. We assume here $k_1 t_1$ and $Q_0t_0$ positive; 
 the other cases can be obtained by reflection or rotation.\label{allowed-regions}}
\end{figure}
Since $r=0$ is not a curvature singularity unless one of the charges is zero, the spacetime can be extended to $r<0$.
The singularities are represented in the $r$-$t$ plane by two hyperbolae having the asymptotes $r=0$ and respectively $t=0$
or $t=-k_1 t_1$; if $k_1\neq 0$ they intersect unless $Q_0 t_0=Q_1 t_1$. To ensure the regularity of the solution we must
require $H_0 H_1 >0$; this corresponds to the area external to the singularities in the $r$-$t$ plane for $t_0 t_1>0$ or to
the area between them if $t_0 t_1<0$ (see figure \ref{allowed-regions}).

The present spacetime satisfies the weak energy condition; to see this, compute the energy-momentum
tensor components $T_{ab}$ for an observer with orthonormal frame
\begin{displaymath}
 e^0=\ma{U}^{-1} dt\,, \quad e^1=\ma{U}\, dr\,, \quad e^2=\ma{U}\, r d\theta\,, \quad
 e^3=\ma{U}\, r \sin\theta d\varphi\,.
\end{displaymath}
One obtains
\begin{eqnarray}
&& \rho^\phi = \ma{F}_1+\ma{F}_2+\ma{F}_3\,, \qquad P_r^\phi = \ma{F}_1+\ma{F}_2-\ma{F}_3\,,
      \qquad P_\Omega^\phi = \ma{F}_1-\ma{F}_2-\ma{F}_3\,, \nonumber \\
&& \rho^{\text{em}} = -P_r^{\text{em}}=P_\Omega^{\text{em}}=\ma{F}_4\,, \qquad
      T_{01}^\phi = -\ma{F}_5\,,
\end{eqnarray}
where $\rho=T_{00}$, $P_r=T_{11}$, $P_\Omega=T_{22}=T_{33}$, the other off-diagonal components are
zero, and
\begin{displaymath}
 \ma{F}_1 = \ma{U}^2   \, \frac{n_0n_1}{16}\left( \frac{1}{t_0H_0}-\frac{1}{t_1H_1} \right)^2\,, \qquad
 \ma{F}_2 = \ma{U}^{-2}\, \frac{n_0n_1}{16r^4}\left( \frac{Q_0}{H_0}-\frac{Q_1}{H_1} \right)^2\,,
\end{displaymath}
\begin{displaymath}
 \ma{F}_3 = \frac{\ma{U}^2}4\left[\left( \frac{n_0}{t_0H_0}-\frac{n_1}{t_1H_1} \right)^2-\frac{n_0}{t_0^2 H_0^2}-
 \frac{n_1}{t_1^2 H_1^2}\right]\,, \qquad
 \ma{F}_4 = \frac{\ma{U}^{-2}}{4r^4}\left( \frac{n_0Q_0^2}{H_0^2} +\frac{n_1Q_1^2}{H_1^2}\right)\,,
\end{displaymath}
\begin{displaymath}
 \ma{F}_5 = \frac{n_0n_1}{8r^2}\left( \frac{1}{t_0H_0}-\frac{1}{t_1H_1} \right)\left(  \frac{Q_0}{H_0}-
 \frac{Q_1}{H_1}\right)\,.
\end{displaymath}
Since $\ma{F}_1$, $\ma{F}_2$, $\ma{F}_4$ and
$\ma{F}_1+\ma{F}_3=\ma{U}^2   \,\frac{3}{16}\left( \frac{n_0}{t_0H_0}+ \frac{n_1}{t_1H_1}\right)^2$ are positive
definite, the energy densities $\rho^\phi$ and $\rho^{\text{em}}$ are positive. Notice also that 
$\ma{F}_4-\ma{F}_2=\ma{U}^{-2}\left( \frac{n_0 Q_0}{H_0}+ \frac{n_1Q_1}{H_1}\right)^2\!/(16r^4)$ is positive
definite and that $\ma{F}_5^2=4\ma{F}_1\ma{F}_2$.

$T^a_{\phantom{a}b}$ can always be diagonalized by changing to a different orthonormal basis.
Its eigenvalues are
\begin{align}
 \hat\rho&=\frac{1}{2}\left( \rho-P_r+\sqrt{(\rho+P_r)^2-4T_{01}^2} \right)\,, \\
 -\hat P_r&=-\frac{1}{2}\left( \rho-P_r+\sqrt{(\rho+P_r)^2-4T_{01}^2} \right)\,, \\
 -\hat P_\Omega&=-P_\Omega\,.
\end{align}
In terms of these the weak energy condition can be stated as
\begin{equation}
 \hat\rho\ge 0\,, \qquad \hat\rho+\hat P_r\ge 0\,, \qquad \hat\rho+\hat P_\Omega\ge 0\,. \label{wec}
\end{equation}
We have
\begin{align}
 &\hat\rho=\ma{F}_3+\ma{F}_4+|\ma{F}_1-\ma{F}_2|\ge (\ma{F}_1+\ma{F}_3)+(\ma{F}_4-\ma{F}_2)\ge0\,, \\
 &\hat\rho+\hat P_r=2|\ma{F}_1-\ma{F}_2|\ge 0\,, \\
 &\hat\rho+\hat P_\Omega=\ma{F}_1+\ma{F}_4+(\ma{F}_4-\ma{F}_2)+|\ma{F}_1-\ma{F}_2|\ge 0\,,
\end{align}
and thus \eqref{wec} holds. Whether the strong and dominant energy conditions are satisfied depends on the values of the 
parameters; it has been shown in particular that the Gibbons-Maeda solution ($t_1\rightarrow\infty$) satisfies the strong 
energy condition if and only if the asymptotic cosmological background does \cite{Maeda:2010ja}, and that the 
Maeda-Ohta-Uzawa solution ($t_1\rightarrow\infty$, $n_0=1$) satisfies the dominant energy condition \cite{Maeda:2009ds}.

The spherical symmetry allows us to covariantly define the circumference radius
$R=|r|\,\ma{U}=|r| H_0^{n_0/4}H_1^{n_1/4}$; it is immediate to see that this radius vanishes on the singularities.
In a spherically symmetric spacetime it is also possible to compute the Misner-Sharp quasilocal
energy \cite{Misner:1964je}, that can be interpreted as the energy inside a closed surface of radius $R$,
\begin{equation}
 m=4\pi R\left(  1+\nabla_\mu R\nabla^\mu R\right)\,,
\end{equation}
where
\begin{equation}
 \nabla_\mu R\nabla^\mu R=-\frac{1}{16}\left[\left( \frac{t n_0}{t_0H_0}+\frac{(t+k_1t_1)n_1}{t_1H_1} \right)^2-r^2H_0^{n_0}H_1^{n_1}\left(\frac{n_0}{t_0H_0} +\frac{n_1}{t_1H_1}\right)^2\right]\,.\label{cond-trapped}
\end{equation}

% \begin{gather}
%  \nabla_\mu R\nabla^\mu R=-\frac{1}{16}\left[(t^2-r^2H_0^{n_0}H_1^{n_1})\left( \frac{n_1}{t_1H_1}+ \frac{n_0}{t_0H_0}\right)^2\right.\nonumber\\%-\frac{1}{16}\left[\left( \frac{n_0 t}{H_0t_0}+\frac{n_1(t+k_1t_1)}{H_1t_1} \right)^2-r^2H_0^{n_0}H_1^{n_1}\left( \frac{n_1}{t_1H_1}+ \frac{n_0}{t_0H_0}\right)^2\right]
%  \left.+k_1 \frac{n_1}{H_1}\left( k_1 \frac{n_1}{H_1} +2t\left( \frac{n_1}{t_1H_1}+ \frac{n_0}{t_0H_0} \right)\right)\right]\,. \label{cond-trapped}
% \end{gather}

Following \cite{Maeda:2009ds,Maeda:2010ja} we can look for trapping horizons \cite{Hayward:1993wb}.
Introducing the Newman-Penrose null tetrads
\begin{eqnarray}
 l&=&\frac{1}{\sqrt{2}}\left( \ma{U}^{-1}dt-\ma{U}dr \right)\,, \nonumber\\
 n&=&\frac{1}{\sqrt{2}}\left( \ma{U}^{-1}dt+\ma{U}dr \right)\,, \\
 m&=&\ma{U}\frac{r}{\sqrt{2}}\left( d\theta+i\sin\theta d\varphi \right)\,, \nonumber
\end{eqnarray}
and the complex conjugate $\bar m$, satisfying $l^\mu n_\mu=1=-m^\mu\bar m_\mu$, the expansions of the outgoing and ingoing 
radial null geodesics are defined by
\begin{equation}
 \theta_+\equiv-2m^{(\mu}\bar m^{\nu)}\nabla_\mu l_\nu\,,\qquad \theta_-\equiv-2m^{(\mu}\bar m^{\nu)}\nabla_\mu n_\nu\,,
\end{equation}
which evaluated explicitly are
\begin{equation}
 \theta_\pm=\frac1{2\sqrt2 r\ma{U}}\left[r\ma{U}^2\left( \frac{n_0}{t_0H_0} +\frac{n_1}{t_1H_1}\right)\pm\left( \frac{t n_0}{t_0H_0}+\frac{(t+k_1t_1)n_1}{t_1H_1} \right)\right].\label{expansions}
\end{equation}
While $\theta_\pm$ are not covariant quantities, their product is; comparing \eqref{cond-trapped} and \eqref{expansions} it
is straightforward to conclude that
\begin{equation}
 \theta_+\theta_-=\frac{2}{R^2}\nabla_\mu R\nabla^\mu R\,.
\end{equation}
A metric sphere is said to be \emph{trapped} or \emph{untrapped} if $\theta_+\theta_->0$ or
$\theta_+\theta_-<0$ respectively, and to be \emph{marginal} if $\theta_+\theta_-=0$. A trapping horizon is
the closure of a hypersurface foliated by marginal surfaces, which means that it occurs when
$\theta_+\theta_-=0$, or equivalently when $\nabla_\mu R$ becomes null.

It is possible to geometrically define on trapping horizons a local surface gravity $k_l$ and the associated
Hawking temperature $T_l=\frac{k_l}{2\pi}$ \cite{Hayward:1997jp,Hayward:2008jq},
\begin{gather}
 k_l\equiv -\frac{1}{2} \tilde\nabla_\mu\tilde\nabla^\mu R\,\big|_{\text{TH}}=\nonumber\\
 -\frac{1}{8 R}\left\{\left(\frac{tn_0t_1H_1+(t+k_1t_1)n_1t_0H_0}{n_0t_1H_1+n_1t_0H_0}  \right)^{\!2}\left[\left( \frac{n_0}{t_0H_0} +\frac{n_1}{t_1H_1}\right)^{\!2}-\left( \frac{n_0}{t_0^2H_0^2} +\frac{n_1}{t_1^2H_1^2}\right)\right]\right.\nonumber\\
  \left.+\left( \frac{t^2 n_0}{t_0^2H_0^2} +\frac{(t+k_1t_1)^2 n_1}{t_1^2H_1^2} \right)-2\left( \frac{t n_0}{t_0H_0} +\frac{(t+k_1t_1)n_1}{t_1H_1} \right)\right\},\label{hayward_sgrav}
\end{gather}
where $\tilde\nabla$ is the covariant derivative associated with the two dimensional metric normal to the
spheres of symmetry. This surface gravity satisfies on the trapping horizons an identity similar to the usual
relation for stationary black holes,
\begin{equation}
 K^\mu\nabla_{[\nu}K_{\mu]}=k_l K_\nu\,,
\end{equation}
where in place of a Killing vector we have the Kodama vector $K\equiv g^{-1}(* dR)$, with $*$ evaluated
with respect to the normal metric. It should be noted however that an observer whose worldline is an integral
curve of $K$ does not measure the temperature $T_l$ near the trapping horizons; the observed temperature
is, to first order, $T=T_l\, C^{-1/2}$, with redshift factor $C=\nabla_\mu R\nabla^\mu R$.

Now if we take $k_1=0$ or equivalently consider the limit $r\to0$, $t\to\infty$ with $rt$ kept finite, 
\eqref{cond-trapped} vanishes for $t^2=r^2H_0^{n_0}H_1^{n_1}$, i.e.,
\eq
t^2r^2 = \left(\frac{tr}{t_0} + Q_0\right)^{n_0}\left(\frac{tr}{t_1} + Q_1\right)^{n_1}\,, \label{cond-trapped-limit}
\feq
or $n_1 t_0 H_0+n_0 t_1H_1=0$ if $t_0t_1<0$. However the latter solution doesn't correspond to a change of sign in
$\theta_+ \theta_-$, so it doesn't identify a trapping horizon. Notice that the solutions of
\eqref{cond-trapped-limit} have constant circumference radius $R$, and since the gradient of $R$
becomes null there, the trapped horizons are null surfaces in the limit $r\to0$, $t\to\infty$ with
$rt$ fixed. In this limit the geometric surface gravity \eqref{hayward_sgrav} simplifies to
\begin{equation}
 k_l=\frac{1}{8 R}\left( \frac{tn_0}{t_0H_0}+ \frac{tn_1}{t_1H_1}\right)\left( 2-\frac{tn_0}{t_0H_0}-\frac{tn_1}{t_1H_1} \right)\label{hayward_sgrav_limit}.
\end{equation}

The identification of event horizons is a nontrivial task for dynamical black holes, since it requires
the knowledge of the entire causal structure of the spacetime. Nevertheless, we can argue as in
\cite{Maeda:2009ds}, and use the fact that the event horizon has to cover the trapped surfaces
provided the outside region of a black hole behaves sufficiently well \cite{Hawking:1973uf}. Since
the spacetime \eqref{eq:mod2_flat_sol} is indeed well-behaved for positive $r$ (as long as we
are outside the forbidden regions in fig.~\ref{allowed-regions}), and the trapping horizons contain
null surfaces \eqref{cond-trapped-limit} in the limit $r\to0$, $t\to\infty$, we shall examine in the following
if these null surfaces are possible candidates for the black hole event horizon. As we said, the limit
$r\to0$, $t\to\infty$ with $rt$ kept finite is equivalent to taking $k_1=0$. In this case,
the metric is invariant under the transformation $t\rightarrow\alpha t,\, r\rightarrow r/\alpha$,
and thus admits the Killing vector $\xi=t\partial_t-r\partial_r$, which is hypersurface orthogonal.
Introducing the coordinates
\begin{gather}
 T=\pm\log|t|+\int^\mathcal{R}\frac{g^2(\mathcal{R})}{\mathcal{R}f(\mathcal{R})}d\mathcal{R}\,, \qquad
 \mathcal{R}=\frac{rt}{Q_0t_0}\,, \\
  \qquad f(\mathcal{R})\equiv (Q_0t_0)^2 \mathcal{R}^2-g^2(\mathcal{R})\,, \qquad g(\mathcal{R})\equiv Q_0^2(\mathcal{R}+1)^\frac{n_0}2\!\left( \frac{t_0}{t_1}\mathcal{R}+\frac{Q_1}{Q_0} \right)^{\!\frac{n_1}2}\,,
\end{gather}
such that $\xi=\partial_T$, the metric can be written in static form as
\begin{equation}
 ds^2=\frac{f(\mathcal{R})}{g(\mathcal{R})} dT^2-(Q_0t_0)^2\frac{g(\mathcal{R})}{f(\mathcal{R})}d\mathcal{R}^2
 -g(\mathcal{R})d\Omega^2\,. \label{metr-static}
\end{equation}
From \eqref{metr-static} it is clear that there are Killing horizons where $f(\mathcal{R})=0$, that is, 
in $(r,t)$ coordinates, $t^2=r^2 H_0^{n_0}H_1^{n_1}$; thus the Killing horizons coincide with the trapping
horizons \eqref{cond-trapped-limit}. As the near-horizon geometry \eqref{metr-static} enjoys the unexpected
symmetry under translations of the time coordinate $T$ (which is not a symmetry of the original
spacetime \eqref{eq:mod2_flat_sol}), our solution \eqref{eq:mod2_flat_sol} provides (like the ones in
\cite{Maeda:2009zi,Maeda:2009ds}) a realization of asymptotic symmetry enhancement at the horizon of
a dynamical black hole. The fact that the horizon does not grow, i.e., the ambient matter does not accrete
onto the black hole, was conjectured in \cite{Nozawa:2010zg} to be related to fake supersymmetry.

Since the spacetime \eqref{metr-static} is static, we can calculate the surface gravity on the horizons which
is given by
\begin{equation}
 k^2=-\frac{1}{2}\nabla_\mu\xi_\nu\nabla^\mu\xi^\nu=\frac{1}{4}\left(\frac{n_0\mathcal{R}}{\mathcal{R}+1}+\frac{n_1 \mathcal{R}}{\mathcal{R}+\frac{Q_1t_1}{Q_0t_0}} -2\right)^2\,, \label{eq:surfacegrav}
\end{equation}
that depends only on $\mathcal{R}$ (or equivalently on $rt$) and where $\mathcal{R}$ is one root of
$f(\mathcal{R})=0$. Note that, contrary to the asymptotically flat case, there is no preferred normalization
for the Killing vector $\xi$ here, and that the surface gravity is sensitive to this norm.
Notice also that in general \eqref{eq:surfacegrav} is nonvanishing. A temperature different from zero
would be in contradiction with supersymmetry, but not with fake supersymmetry: Following the
explanation in \cite{Kostelecky:1995ei}, consider a black hole with temperature $T$. A spinor
in the Euclidean section must then be antiperiodic under translation of the Euclidean time through a
period $\beta=1/T$. Supersymmetry implies the existence of a spinor field solving the Killing spinor
equation, and this spinor must be periodic to give a regular solution. Both requirements are compatible
only if the period is infinite, or equivalently when the temperature vanishes. Now, in fake supergravity,
there are no fermions whose variation under a putative fake supersymmetry transformation is associated
to the fake Killing spinor equation. The latter is just an auxiliary construction, which implies (under
certain conditions) the second order field equations. Thus, the above contradiction for nonzero
temperature does not arise.

Rewriting \eqref{hayward_sgrav_limit} in static coordinates,
\begin{equation}
 k_l=\frac{1}{8 \sqrt{|Q_0t_0\ma{R}|}}\left( \frac{n_0\ma{R}}{\ma{R}-1}+\frac{n_1\ma{R}}{\ma{R}-\frac{Q_1t_1}{Q_0t_0}} \right)\left( 2-\frac{n_0\ma{R}}{\ma{R}-1}-\frac{n_1\ma{R}}{\ma{R}-\frac{Q_1t_1}{Q_0t_0}} \right)\,,
\end{equation}
we see that it agrees with \eqref{eq:surfacegrav} up to a normalization factor constant over each Killing horizon. This is the same factor that ties the Kodama vector $K$ to the Killing vector $\xi$ on the horizons,
\begin{equation}
 K|_{\text{KH}}=\pm\frac{1}{4 \sqrt{|Q_0t_0\ma{R}|}}\left( \frac{n_0\ma{R}}{\ma{R}-1}+\frac{n_1\ma{R}}{\ma{R}-\frac{Q_1t_1}{Q_0t_0}} \right)\xi\,.
\end{equation}

The horizon condition can be rewritten as
\begin{equation}
 |\mathcal{R}|=a |\mathcal{R}+1|^{\frac{n_0}{2}} |\mathcal{R}+b|^{\frac{n_1}{2}},\qquad a\equiv\left|\frac{Q_0}{t_0}\right|\left| \frac{t_0}{t_1} \right|^{\frac{n_1}{2}},\quad b\equiv\frac{Q_1 t_1}{Q_0 t_0}\,.
\end{equation}
If $t_0 t_1>0$ the accessible regions of spacetime are $\mathcal{R}>\text{max}(-1,-b)$ and
$\mathcal{R}<\text{min}(-1,-b)$. 

We see that for $b>1$ there are always exactly two horizons for negative $\mathcal{R}$, one for
$\mathcal{R}<-b$ and one for $-1<\mathcal{R}<0$; however only one of these is accessible since they are
located in disconnected regions of the spacetime. For $a\geq1/4$ one has
$\mathcal{R}\leq a(\mathcal{R}+1)^2<a (\mathcal{R}+1)^{n_0/2} (\mathcal{R}+b)^{n_1/2}$ for every positive 
$\mathcal{R}$ and consequently there are no other horizons. On the other hand if $a\leq1/(4 \,b)$ there is an interval for which $\mathcal{R}\geq a(\mathcal{R}+b)^2>a (\mathcal{R}+1)^{n_0/2} (\mathcal{R}+b)^{n_1/2}$
and there are thus two distinct horizons for positive $\mathcal{R}$. For intermediate values of $a$ there can be
zero or two, possibly coincident, horizons for positive $\mathcal{R}$ depending on the value of the parameters.
$b=1$ corresponds to the single-centered Kastor-Traschen solution, or Reissner-Nordstr\"om-de~Sitter with
mass equal to the charge, the extremal case corresponding to $a=1/4$. We can then identify the three
horizons in the $\mathcal{R}>-1$ region as respectively inner and outer black hole horizons and cosmological
horizon.

For $b\le -1$ there is always one horizon in the region $\mathcal{R}>-b$ and at least one, at most three
horizons for $\mathcal{R}<-1$. In this case $\mathcal{R}=0$ is not accessible. 

For $b=0$, corresponding to a black hole charged under only one of the gauge fields, there is a solution in
$\mathcal{R}=0$ which is not a horizon since it is coincident with a singularity; depending on the value of the
parameters there can be zero, one or two horizons for $\mathcal{R}>0$. 
In the region  $\mathcal{R}<-1$ there is always a single horizon.

If $t_0 t_1<0$ the accessible region is given by the values of $\mathcal{R}$ between $-1$ and $-b$.
For $b>1$ there can be zero, one or two, possibly coincident, horizons; for $b<-1$ there are always two
horizons, one with negative and one with positive $\mathcal{R}$. For $b=0$ there is again a solution in
$\mathcal{R}=0$ coincident with a singularity; depending on the value of the parameters there can be zero,
one or two additional solutions corresponding to horizons.

With the choice of coordinates we made, the radial null geodesic equations simplify to
\begin{equation}
 \ddot T +2\,{\Gamma^{\scriptscriptstyle{T}}}_{\scriptscriptstyle{T\mathcal{R}}}\dot T \,\dot{\mathcal{R}}=0\,,
 \qquad \ddot{\mathcal{R}}=0\,,
\end{equation}
which means that $\mathcal{R}$ is an affine parameter for the radial null geodesics and consequently all
horizons and singularities are reached within a finite value of the affine parameter. From the null condition 
$d\mathcal{R}=\pm f(\mathcal{R})dT/(Q_0 t_0 g(\mathcal{R}))$ we obtain the 
expressions for the radial null geodesics in the near-horizon and near-singularity limits,
\begin{equation}
 \begin{array}{ll}
  \mathcal{R}\sim\mathcal{R}_{\text{hor}}&:\qquad T=\pm\frac1{2k}\log|\mathcal{R}-\mathcal{R}_{\text{hor}}|
  +c_1\,, \\
  \mathcal{R}\sim -1&:\qquad T=\pm 2\frac{Q_0}{t_0}\left( \frac{Q_1}{Q_0}-\frac{t_0}{t_1} \right)^{\frac{n_1}2}
  \frac{\left( \mathcal{R}+1 \right)^{1+\frac{n_0}2}}{2+n_0}+c_2\,, \\
  \mathcal{R}\sim -\frac{Q_1t_1}{Q_0t_0}&:\qquad T=\pm 2\frac{Q_1}{t_1}\left( \frac{Q_0}{Q_1}-\frac{t_1}{t_0}
  \right)^{\frac{n_0}2}\frac{\left( \frac{Q_0t_0}{Q_1t_1}\mathcal{R}+1 \right)^{1+\frac{n_1}2}}{2+n_1}+c_3\,,
 \end{array}
\end{equation}
where $c_i$ are constants and $k$ is the surface gravity (\ref{eq:surfacegrav}).

\section{Alternative model}
\label{alternate-prepot}

In section \ref{sec:mod2} we considered the truncation $\mathfrak{Im}(Z)=0$; we could also have taken 
$\mathfrak{Re}(Z)=0$, but this choice is not consistent for every value of $n$ with the prepotential we had
there. Here we consider a slightly modified prepotential,
\eq
\mathcal{F}(\chi)=\frac{i^{n-1}}{4(1-n)}(\chi^0)^n(\chi^1)^{2-n}\,, \label{prepot-2}
\feq
with $n\neq 0,1,2$, that leads to consistent results with the truncation $\mathfrak{Re}(Z)=0$ (but not with
$\mathfrak{Im}(Z)=0$). The model \eqref{prepot-2} is of course related to the one of section \ref{sec:mod2}
by a complex rescaling of the $\chi^{\Lambda}$, and thus the truncations considered here and in the
preceding section are actually two different truncations of the same model.

From (\ref{eq:sympcond2}), taking $\left|\chi^0\right|=1$ we obtain the K\"ahler potential
\begin{equation}
 e^{-\mathcal{K}}=\frac{i^n}{4(1-n)} Z^{1-n}[nZ+(2-n)\bar Z ] + \text{c.c.}\,,
\end{equation}
and, imposing $\mathfrak{Re}(Z)=0$, the K\"ahler metric
\begin{equation}
 \mathcal{G}=\left.\partial_Z\partial_{\bar Z}\mathcal{K}\right|_{\mathfrak{Re}(Z)=0}=-\frac{n(2-n)}{4}
 \mathfrak{Im}(Z)^{-2}=\frac{n_0n_1}{(n_1-n_0)^2}e^{\frac{n_0-n_1}{2}\phi}\,,
\end{equation}
with $n_0\equiv -\frac{2n}{1-n}$, $n_1\equiv \frac{2(2-n)}{1-n}=4-n_0$, $\phi\equiv\frac{4}{n_1-n_0}\log\mathfrak{Im}(Z)$.

From equation (\ref{eq:nmatrix}) one obtains the vectors' kinetic matrix
\begin{equation}
 \mathcal{N}=-\frac{i}{8}\left(\begin{array}{cc}
                    n_0 e^{\frac{n_1}{2}\phi} & 0                 \\
                    0 & n_1 e^{\frac{n_0}{2}\phi}
                   \end{array}
\right),
\end{equation}
while (\ref{eq:scalar_pot}) leads to the scalar potential
\begin{align}
 V=\frac12\left[\frac{n_0(n_0-1)}{t_0^2}e^{-\frac{n_1}{2}\phi}+\frac{n_1(n_1-1)}{t_1^2}e^{-\frac{n_0}{2}\phi}\right]\,,
\end{align}
where we defined as before $t_\Lambda\equiv-\frac{n_\Lambda}{2gC_\Lambda}$.

Substituting in eq. (\ref{eq:general_lagrangian}) we have
\begin{align}
 e^{-1}\mathcal{L}=&R+\frac{n_0n_1}{8}\partial_\mu\phi\partial^\mu\phi-\frac{n_0}{4}e^{\frac{n_1}{2}\phi} F^0_{\mu\nu}F^{0\mu\nu}-\frac{n_1}{4}e^{\frac{n_0}{2}\phi} F^1_{\mu\nu}F^{1\mu\nu}\nonumber\\
                   &-\frac{1}{2}\left[\frac{n_0(n_0-1)}{t_0^2}e^{-\frac{n_1}{2}\phi}+\frac{n_1(n_1-1)}{t_1^2}e^{-\frac{n_0}{2}\phi}\right]\,,
\end{align}
which differs from the Lagrangian obtained in the previous section only by a sign in front of $n_0$ in the
exponents and the absence of the cross term in the potential. To avoid ghost fields in the action we must
restrict $n_0$ and $n_1$ to positive values, which corresponds to have in the prepotential either $n<0$ or
$n>2$.

(\ref{eq:risections}), together with $\mathfrak{Re}(Z)=0$, leads to
\begin{gather}
 \mathcal{I}_0=-\frac{n_0}{8} e^\phi \mathcal{I}^1\,, \qquad \mathcal{I}_1=\frac{n_1}{8}e^\phi
 \mathcal{I}^0\,, \nonumber\\
 \mathcal{R}^0=e^{\frac{n_0-n_1}4\phi}\mathcal{I}^1\,, \qquad \mathcal{R}^1=-e^{\frac{n_1-n_0}4\phi}
 \mathcal{I}^0\,, \nonumber\\
 \mathcal{R}_0=\frac{n_0}8 e^{\frac{n_1}2\phi}\mathcal{I}^0\,, \qquad \mathcal{R}_1=\frac{n_1}8
 e^{\frac{n_0}2\phi}\mathcal{I}^1\,, \label{eq:mod1ir}
\end{gather}
as well as
\begin{equation}
 \frac{1}{2|X|^2}=\langle\mathcal{R}|\mathcal{I}\rangle=\frac{1}{2}\left[e^{\frac{n_1}{2}\phi}(\mathcal{I}^0)^2+e^{\frac{n_0}{2}\phi}(\mathcal{I}^1)^2\right]\,. \label{eq:mod1sympprod}
\end{equation}

From (\ref{eq:mod1ir}) and (\ref{eq:dertaui}) we see that, since we exclude the case $C_0=C_1=0$, 
$\mathcal{I}^1=0$ is equivalent to $C_0=0$, $\mathcal{I}^0=0$ is equivalent to $C_1=0$, and
$C_1\mathcal{I}^1=-\frac{n_1}{n_0}C_0\mathcal{I}^0$.

\subsection{Construction of the solution}

As before we take
\begin{equation}
 W^x_m=\delta^x_m\,, \qquad C_\Lambda\tilde A^\Lambda=C_\Lambda \mathcal{I}^\Lambda=0\,.
\end{equation}
Since $C_1\mathcal{I}^1=-\frac{n_1}{n_0}C_0\mathcal{I}^0$, $C_\Lambda \mathcal{I}^\Lambda=0$ with
$n_0\neq 2$\footnote{For $n_0=2$ ($n\rightarrow\pm\infty$) we could take both $C_0, C_1\neq 0$
(equivalently,  $\mathcal{I}^0,\mathcal{I}^1\neq 0$); however this would lead to exactly the same solution we
obtain here, with just a field redefinition.} implies $C_0\mathcal{I}^0=0$. One has thus either
$C_0=\mathcal{I}^1=0$ or $C_1=\mathcal{I}^0=0$. We will consider just the first case since the
second can be obtained simply by exchanging $0$ and $1$ indices. We have thus
\begin{equation}
 C_0=0\,, \qquad \mathcal{I}^1=\mathcal{I}_0=\mathcal{R}^0=\mathcal{R}_1=0\,, \label{eq:mod1flatzerosections}
\end{equation}
and from $C_\Lambda\tilde A^\Lambda=0$, taking into account that $C_1\neq 0$, 
\begin{equation}
 \tilde A^1=0\,.
\end{equation}
Eq.~(\ref{eq:bogo}) yields
\begin{equation}
 \tilde{F}^0_{mn}=-\frac{1}{\sqrt{2}}\varepsilon_{mnp}\partial_p \mathcal{I}^0\,,
\end{equation}
and from the Bianchi identity $d\tilde F^0=0$ we obtain
\begin{equation}
 \partial_p\partial_p\mathcal{I}^0=0\qquad\Rightarrow\qquad \mathcal{I}^0=\sqrt{2}\mathcal{H}_0\,,
\end{equation}
where $\mathcal{H}_0$ is a generic harmonic function of the spatial coordinates.

Using (\ref{eq:mod1flatzerosections}) and $C_\Lambda \tilde{A}^\Lambda=0$,  from eq. (\ref{eq:omegai2})
we conclude as before $\tilde\omega=df$, where $f$ is a generic function of the spatial coordinates.
(\ref{eq:omegai1}) implies then
\begin{equation}
         \partial_p\partial_p\left(\tilde{\mathcal{I}}_1+\frac{g C_1}{2\sqrt{2}}f\right)=0 \qquad\Rightarrow\qquad\tilde{\mathcal{I}}_1=\frac{n_1}{4\sqrt{2}}(f/t_1+\mathcal{H}_1)\,,
\end{equation}
where $\mathcal{H}_1$ is another harmonic function of the spatial coordinates.
Using (\ref{eq:dertaui}) and (\ref{eq:mod1ir}) one gets
\begin{equation}
 \mathcal{I}_1=\frac{n_1}{4\sqrt2}((\tau+f)/t_1+\mathcal{H}_1)\,, \qquad
 e^\phi=\frac{(f+\tau)/t_1+\mathcal{H}_1}{\mathcal{H}_0}\,,
\end{equation}
and from (\ref{eq:mod1sympprod}) one computes
\begin{equation}
 \frac{1}{2|X|^2}=\left(\frac{\tau+f}{t_1} + \mathcal{H}_1\right)^\frac{n_1}{2}\mathcal{H}_0^\frac{n_0}{2}.
\end{equation}
Eliminating $f$ by introducing the new time coordinate $t=\tau+f$, the solution can be written as
\begin{gather}
ds^2=\mathcal{U}^{-2}dt^2- \mathcal{U}^2 d\vec y^{\,2}\,, \nonumber\\
F^0=-\frac12\varepsilon_{mnp}\partial_p\mathcal{H}_0\, dy^m \wedge dy^n\,, \qquad
A^1=\left(\frac{t}{t_1} + \mathcal{H}_1\right)^{-1}dt\,, \label{eq:mod1_flat_sol}\\
\phi=\ln\left(\frac{t/t_1+\mathcal{H}_1}{\mathcal{H}_0}\right)\,, \nonumber
\end{gather}
with
\begin{equation}
 \mathcal{U}\equiv\left(\frac{t}{t_1} + \mathcal{H}_1\right)^\frac{n_1}{4}\mathcal{H}_0^\frac{n_0}4.
\end{equation}
This is, with the right choice for $\mathcal{H}_0$ and $\mathcal{H}_1$, the spacetime found in
\cite{Gibbons:2009dr} and discussed further in \cite{Maeda:2010ja}; however in this case, instead of having
two gauge fields in an electric configuration, one of them is magnetic due to the different sign in the
exponent of its scalar coupling. In other words, one of the field strengths in the Gibbons-Maeda solution
is dualized here.

\section{Final remarks}
\label{final}

Let us conclude our paper with the following suggestions for possible extensions
and questions for future work:
\begin{itemize}
\item Add rotation. This is under investigation \cite{Chimento:2013xx}.
\item Construct the corresponding `nonextremal' solution (i.e., the one that does not admit fake
Killing spinors), which might be of astrophysical relevance.
\item Do our solutions allow to study dynamical processes like black hole collisions, similar to what was
done in \cite{Kastor:1992nn,Brill:1993tm} for the Kastor-Traschen spacetime?
\item Does the attractor mechanism \cite{Ferrara:1995ih,Strominger:1996kf,Ferrara:1996dd,Ferrara:1996um,Ferrara:1997tw} continue to work in the time-dependent case? This issue has not been addressed
in the literature so far.
\item One may consider more general Gauduchon-Tod base spaces and/or more complicated
prepotentials in the construction of \cite{Meessen:2009ma}, and see whether this leads to physically
interesting solutions.
\end{itemize}

\end{document}